# A Novel Simplified Swarm Optimization for Generalized Reliability Redundancy Allocation Problem

Zhenyao Liu[1], Jen-Hsuan Chen[1], Shi-Yi Tan[1], Wei-Chang Yeh[1]

[1]Integration and Collaboration Laboratory, Department of Industrial Engineering and Engineering Management, National Tsing Hua University, Hsinchu 30013, Taiwan

**Abstract**

Network systems are commonly used in various fields, such as power grid, Internet of Things (IoT), and gas networks. Reliability redundancy allocation problem (RRAP) is a well-known reliability design tool, which needs to be developed when the system is extended from the series-parallel structure to a more general network structure. Therefore, this study proposes a novel RRAP called General RRAP (GRRAP) to be applied to network systems. The Binary Addition Tree Algorithm (BAT) is used to solve the network reliability. Since GRRAP is an NP-hard problem, a new algorithm called Binary-addition simplified swarm optimization (BSSO) is also proposed in this study. BSSO combines the accuracy of the BAT with the efficiency of SSO, which can effectively reduce the solution space and speed up the time to find high-quality solutions. The experimental results show that BSSO outperforms three well-known algorithms, Genetic Algorithm (GA), Particle Swarm Optimization (PSO), and Swarm Optimization (SSO), on six network benchmarks.

## 1. INTRODUCTION

The reliability of network systems has long been a focus of research in various fields, such as computing and communications, power transmission, distribution and transportation, and the IoT [1-7]. Many studies have shown that the systems in engineering, industrial and scientific applications can benefit from improving reliability from the initial phases of design [8-12].

Reliability design is an important tool in system design and management. There are two main approaches to improve reliability: improving component reliability and providing redundant components [10, 12-14]. These two approaches can create different types of reliability design problems. The first is the Redundancy allocation problem (RAP), which focuses on optimizing redundant components as a decision variable in the case of determining component reliability.

Conventional RRAP subsystems are mostly connected in series or series-parallel, and the components in the subsystems are connected in parallel. Let $m$ be the number of subsystems in the series, $n_i$ be the number of components that can be used in parallel in subsystem $i$, $r_i$ be the reliability of each component used in subsystem $i$. Since the components in the subsystem are in parallel, the $i$th subsystem reliability $R_i$ is calculated as follows:

$$R_i = 1 - (1 - r_i)^{n_i} \qquad (1)$$

As mentioned earlier, most of the RRAP connect subsystems in series or series-parallel. However, many systems are more complex general network systems, so RRAP subsystem connections need to be developed more generally. Yeh proposed a generalized redundancy allocation problem (GRAP) [15], which is an extension of RAP to the general network system. In the GRAP, subsystems are not limited to series or series-parallel connections. Each subsystem can be connected to any subsystem in the system. In this study, the activity on arc (AOA) network is used as the system structure. However, to the best of our knowledge, no studies have discussed the generalized RRAP (GRRAP), which extends traditional RRAP to general subsystem structures. Fig. 1 exemplifies a GRRAP configuration with six subsystems.

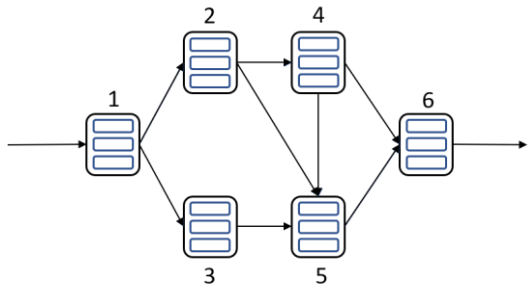

**Fig. 1.** A GRRAP configuration

The RRAP is a well-known NP-hard problem [13], whose computational resource requirements will grow exponentially as the number of components in the system increases. Therefore, it is difficult to find a solution in a reasonable time. To reduce its computational burden, many studies have been devoted to developing soft computing methods such as artificial bee colony algorithm (ABC) [16], genetic algorithm (GA) [17, 18], immune algorithm (IAs) [19], simplified swarm optimization (SSO) [20, 21], combining both PSO and SSO (PSSO) [22].

When RRAP is extended to GRRAP, the computation complexity of GRRAP is more difficult because the network reliability calculation is also an NP-hard problem. Therefore, this paper proposed a new soft computing method called Binary-addition Simplified Swarm Optimization (BSSO), which combines the Binary Addition Tree algorithm (BAT) proposed by Yeh [23], and the update mechanism of SSO [24]. The features and advantages of this algorithm are that it is simple but can produce high-quality solutions efficiently.

This paper is organized as follows, Section 2 provides the definition of the proposed GRRAP. Section 3 presents the research methods used to solve the problem, including the network reliability calculation, multi-state BAT, and two different SSO update mechanisms for updating discrete and continuous variables respectively. Section 4 explains the proposed BSSO method for solving the GRRAP, including the solution representation, a new SSO-based update mechanism combining multi-state BAT, a new penalty function that remove weight and volume constraint, and finally the flow chart. Section 5 presents a designed experiment according to orthogonal array test $L_8(2^4)$ to get the best combination of update mechanisms. To verify the performance of BSSO, further comparison with GA, PSO, and SSO was made. Finally, the conclusion and discussion are given in Section 6.

## 2. Problem description of the proposed GRRAP

The GRRAP model proposed in this study is an active RRAP that extends to general network systems, allowing subsystems to be connected to any subsystem in the system without being limited to serial connections. Unlike GRAP, which is presented in an AOA network [15], this study uses the more intuitive AON network, which transforms each subsystem into a node in the network. The reliability of each subsystem depends on the reliability of the parallel components in the subsystem, not all of which are reliable. The network reliability problem with unreliable nodes and reliable arcs can be solved by using the network reliability solution method [23].

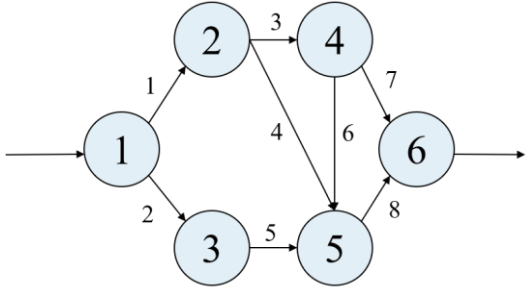

**Fig. 2.** AON binary-state network transformed from Fig. 1.

### 2.1. Proposed GRRAP

The GRRAP proposed in this study is a nonlinear mix-integer programming problem that requires determining the number of components and component reliability in each subsystem to maximize the system reliability. However, GRRAP only extends the system into the general network, so the problem assumptions are almost identical to RRAP. The only difference is that the system reliability is calculated using the network reliability calculation method.

The mathematical model of GRRAP is as follow:

$$\text{Max} \quad R_s(N, R) \qquad (2)$$

s.t.

$$g_c(N, R) = \sum_{i=1}^{N_{var}} \alpha_i (-1000/\ln r_i)^{\beta_i}(n_i + \exp(n_i/4)) \leq C_{ub} \qquad (3)$$

$$g_v(N, R) = \sum_{i=1}^{N_{var}} w_i v_i^2 n_i^2 \leq V_{ub} \qquad (4)$$

$$g_w(N, R) = \sum_{i=1}^{N_{var}} w_i n_i \exp(n_i/4) \leq W_{ub} \qquad (5)$$

$$n_{lb} \leq n = (n_1, n_2, \ldots, n_{N_{var}}) \leq n_{ub} \qquad (6)$$

$$r_{lb} \leq r = (r_1, r_2, \ldots, r_{N_{var}}) \leq r_{ub} \qquad (7)$$

The objective function provided in Eq. (2) is set to maximize $R_s(N, R)$. Eqs. (3)-(5) require that the final solution satisfy three conditions: the total cost $g_c(N, R)$, volume $g_v(N, R)$, and weight $g_w(N, R)$ must be less than or equal to the predefined limits $C_{ub}$, $V_{ub}$, and $W_{ub}$, respectively. Eq. (6) represents the number of components that must be between the upper and lower bounds, while Eq. (7) regulates the range of component reliability.

### 2.2. GRRAP example description

The following example further explicates the GRRAP in Fig. 2 to demonstrate how to calculate the reliability of each subsystem, cost, volume, and weight if a solution $X = (N, R)$ is given. Assume that the system parameters of the network system in Fig.2 are shown in Table 1, and calculate the subsystem reliability with a solution $X = (4, 2, 2, 2, 2, 2, 3, 0.8168, 0.8534, 0.8554, 0.8740, 0.8288, 0.8781)$. The first six coordinates of the vector represent the number of components and the last six coordinates are the corresponding component reliability.

**Table 1** Information for the components in Fig. 2.

| i | $10^5\alpha_i$ | $\beta_i$ | $w_iv_i^2$ | $w_i$ | V | C | W |
|---|---|---|---|---|---|---|---|
| 1 | 2.5 | 1.5 | 2 | 3.5 | 220 | 210 | 120 |
| 2 | 1.45 | 1.5 | 4 | 4.0 | | | |
| 3 | 0.541 | 1.5 | 5 | 4.0 | | | |
| 4 | 0.541 | 1.5 | 8 | 3.5 | | | |
| 5 | 2.1 | 1.5 | 4 | 4.5 | | | |
| 6 | 2.1 | 1.5 | 4 | 4.5 | | | |

* $i$: the $i_{th}$ subsystem

**Table 2** Subsystem reliability for Fig.2.

| i | $R_i$ | | $C_i$ | $W_i$ | $V_i$ |
|---|---|---|---|---|---|
| 1 | $1-(1-0.8168)^4$ 0.99887 | = | 58.3456 | 38.0559 | 32 |
| 2 | $1-(1-0.8534)^2$ 0.97851 | = | 26.5066 | 13.1898 | 16 |
| 3 | $1-(1-0.8554)^2$ 0.97909 | = | 10.1129 | 13.1898 | 20 |
| 4 | $1-(1-0.8740)^2$ 0.98412 | = | 12.6301 | 11.5410 | 32 |
| 5 | $1-(1-0.8288)^2$ 0.97069 | = | 29.7782 | 14.8385 | 16 |
| 6 | $1-(1-0.8781)^3$ 0.99819 | = | 72.5013 | 28.5795 | 36 |
| SUM | | | 209.8747 | 119.3945 | 152 |

The reliability calculation for each subsystem can be obtained according to Eq. (1) like column 2 in Table 2. After calculating the reliability of each subsystem, we can further calculate the reliability of the whole system. The calculation of system reliability will be explained in later section. The remaining cost, volume, and weight calculations are described below and column 3-5 in Table 2.

$g_c(N, R): \sum_{i=1}^{6} \alpha_i(-1000/\ln r_i)^{\beta_i}(n_i + \exp(n_i/4))$ =209.8747

$g_v(N, R): \sum_{i=1}^{6} w_iv_i^2n_i^2$ =152

$g_w(N, R): \sum_{i=1}^{6} w_in_i \exp\left(\frac{n_i}{4}\right)$ =119.3954

## 3. Overview of the network reliability methods, multi-state BAT, and SSO

### 3.1. Network reliability methods

Before demonstrating how to calculate the network reliability calculation, it is necessary to introduce the BAT algorithm, a new heuristic search method proposed by Yeh [23]. BAT is easy to understand, encode, and can be customized according to the problem requirements. Therefore, BAT has been used in many areas of research, such as wildfire propagation [25], computer viruses [26], and network reliability [23, 27, 28]. It has also evolved from the traditional binary state BAT to multi-state BAT [29].

BAT is one of the direct algorithms in the network reliability algorithm. It has been proved in past studies that BAT is faster than the MP algorithms and MC algorithms from the viewpoint of the time complexity [23]. It can find all possible state vectors of the network by binary addition computing, and during the process uses Path-based Layered Search Algorithm (PLSA) and other methods to remove the state vectors that can't be connected. The network reliability is obtained by calculating the occurrence probability and the summed probability of the connection state vectors. The pseudocode of BAT is described below:

| **Pseudocode for Traditional Binary-Addition Tree** | |
|---|---|
| **Input:** | $G(V,E)$, $V$ = node set, $E$ = arc set |
| **Output:** | All possible node state vectors without duplications |
| **Step 1** | Let $SUM = 0$, $k = 1$, and let $X_1 = X$ be a zero vector with $m$ coordinates |
| **Step 2** | Let $i = m$ |
| **Step 3** | If $X(v_i) = 0$, let $X(v_i) = 1$, $k = k+1$, $X_k = X$, $SUM = SUM + 1$, and proceed to **Step 5** |
| **Step 4** | Let $X(v_i) = 0$, if $i > 1$, $i = i - 1$, and proceed to **Step 3**。 |
| **Step 5** | If $SUM = m$, halt and $X_1, X_2, \dots, X_k$ are all possible node state vectors. Otherwise, proceed to **STEP 2** |

And we continue to demonstrate the network reliability calculation of Fig 2 using the subsystem reliability in Table 2. Originally, the BAT is used to find $2^6$ number of node state vectors, because the system has 6 subsystems. However, since the network is a two-terminals AON network, the basic connection requirement is that the node states of the two terminals must be 1. Therefore, only 4 nodes between are required to be found by BAT. It means the number of state vector are reduced from $2^6$ to $2^4$. All node state vectors are listed in Table 3.

**Table 3** All node state vector found by BAT.

| $i$ | $X_i$ | $i$ | $X_i$ |
|---|---|---|---|
| 1 | (1, **0, 0, 0, 0**, 1) | 9 | (1, **1, 0, 0, 0**, 1) |
| 2 | (1, **0, 0, 0, 1**, 1) | 10 | (1, **1, 0, 0, 1**, 1) |
| 3 | (1, **0, 0, 1, 0**, 1) | 11 | (1, **1, 0, 1, 0**, 1) |
| 4 | (1, **0, 0, 1, 1**, 1) | 12 | (1, **1, 0, 1, 1**, 1) |
| 5 | (1, **0, 1, 0, 0**, 1) | 13 | (1, **1, 1, 0, 0**, 1) |
| 6 | (1, **0, 1, 0, 1**, 1) | 14 | (1, **1, 1, 0, 1**, 1) |
| 7 | (1, **0, 1, 1, 0**, 1) | 15 | (1, **1, 1, 1, 0**, 1) |
| 8 | (1, **0, 1, 1, 1**, 1) | 16 | (1, **1, 1, 1, 1**, 1) |

\* The states of vectors shown in blue are obtained by BAT

Before using PLSA to check whether each state vector is connectable, it is possible to compare the sum of node state vectors with $np$ to reduce the number of nodes to be examined. $np$ is defined here as the minimum number of nodes required to create a connectable path, for example, np is 4 in Fig 2, so all state vectors that add up to less than 4 can be removed

Then we use the PLSA to do remaining checking process. The PLSA is a search algorithm, which can search the network layer by layer through the connectable nodes and check whether the network can connect the two terminals. Here we demonstrate the PLSA verification process with the state vector $X_7 = (1,0,1,1,0,1)$. From Table 4, the node of the first layer is node 1, so $L_1 = \{1\}$, and the nodes of the second layer $L_2$ have node 3, so $L_2 = \{3\}$. Node 3 has no nodes that can be connected to it, so the layer $L_{(2+1)}$ is the empty set $\emptyset$, and finally, the set of nodes $V^*$ does not contain the sink node. It is verified that the subnetwork $X_7$ is not a connected state vector. The red letters in Table 5 are the state vectors removed by the PLSA.

**Table 4** Process from the PLSA for $X_7 = (1,0,1,1,0,1)$.

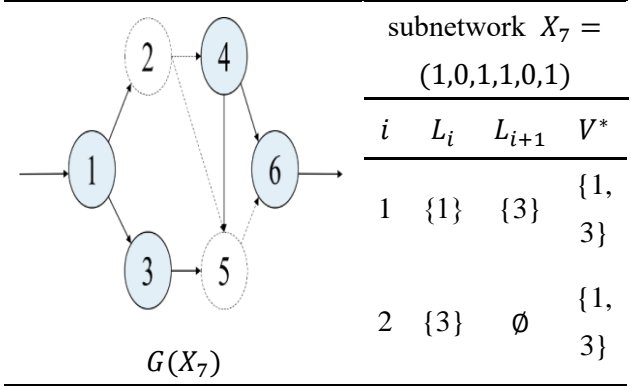

| | subnetwork $X_7 = (1,0,1,1,0,1)$ | | |
|---|---|---|---|
| $i$ | $L_i$ | $L_{i+1}$ | $V^*$ |
| 1 | {1} | {3} | {1, 3} |
| 2 | {3} | ∅ | {1, 3} |

$G(X_7)$

Therefore, the probability of each state vector can be calculated by substituting the node reliability, as in Eq. (8). The sum of these probabilities is the network reliability, as in Eq. (9). The detailed calculation procedure is shown in Table 6.

**Table 5** All the state vectors that can be connected.

| $i$ | $X_i$ | $i$ | $X_i$ |
|---|---|---|---|
| 1 | (1, 0, 1, 0, 1, 1) | 5 | (1, 1, 0, 1, 1, 1) |
| 2 | (1, 0, 1, 1, 1, 1) | 6 | (1, 1, 1, 0, 1, 1) |
| 3 | (1, 1, 0, 0, 1, 1) | 7 | (1, 1, 1, 1, 0, 1) |
| 4 | (1, 1, 0, 1, 0, 1) | 8 | (1, 1, 1, 1, 1, 1) |

$$Pr(X_i) = \prod_{j=1}^{m} Pr(X_i(R_i)) \quad (8)$$
$$R_s = \sum_{\forall X} Pr(X) \quad (9)$$

**Table 6** Final network reliability calculation.

| $i$ | $X_i$ | $Pr(X_i(R_i))$ | $Pr(X_i)$ |
|---|---|---|---|
| 1 | (1, 0, 1, 0, 1, 1) | 0.99887 × (1- 0.97851) × 0.97909 × (1- 0.98412) × 0.97069 × 0.99819 | 0.00032 |
| 2 | (1, 0, 1, 1, 1, 1) | 0.99887 × (1- 0.97851) ×0.97909 ×0.98412 × 0.97069 ×0.99819 | 0.02004 |
| 3 | (1, 1, 0, 0, 1, 1) | 0.99887 ×0.97851 × (1- 0.97909) × (1- 0.98412) × 0.97069 ×0.99819 | 0.00031 |
| 4 | (1, 1, 0, 1, 0, 1) | 0.99887 ×0.97851 × (1- 0.97909) ×0.98412 × (1- 0.97069) ×0.99819 | 0.00059 |
| 5 | (1, 1, 0, 1, 1, 1) | 0.99887 ×0.97851 × (1- 0.97909) ×0.98412 ×0.97069 ×0.99819 | 0.01949 |
| 6 | (1, 1, 1, 0, 1, 1) | 0.99887 ×0.97851 ×0.97909 × (1- 0.98412) ×0.97069 ×0.99819 | 0.01472 |
| 7 | (1, 1, 1, 1, 0, 1) | 0.99887 ×0.97851 ×0.97909 ×0.98412 × (1- 0.97069) ×0.99819 | 0.02755 |
| 8 | (1, 1, 1, 1, 1, 1) | 0.99887 ×0.97851 ×0.97909 ×0.98412 ×0.97069 ×0.99819 | 0.91251 |
| **SUM** | | | **0.99553** |

At the end, the network reliability $R_s$= 0.00032 + 0.02004 + 0.00031 + 0.00059 + 0.01949 + 0.01472 + 0.02755 + 0.91251 = 0.99553. The above is the complete network

reliability calculation process when a solution $X$ is calculated to obtain the subsystem reliability.

## 3.2. Multi-state BAT

In BSSO, the integer variable represents the number of components in the node, and it is necessary to use multi-state BAT to find out all the state vectors $X = (x_1, x_2, ..., x_m)$, each state vector represent an integer combination. The traditional multi-state BAT algorithm is proposed by Yeh [29]. As the number of subsystems and components increases, the number of variable combinations that BAT needs to find will become very large. Therefore, this study proposes an improved multi-state BAT shown as follow:

**Pseudocode for Improved Multi-state Binary-Addition Tree Algorithm**

| | |
|---|---|
| **Input:** | Upper and lower bound set $(UB, LB)$ of integer variable, weight limit, and volume limit |
| **Output:** | A set $Comb$ that contains all combinations of integer variables that satisfy the constraints |
| **Step 1:** | Let $i = m$, $x_k = lb_k$ for $k = 0, 1, 2, ..., m$ |
| **Step 2:** | If $x_i < ub_i$, let $x_i = x_i + 1$, $i = m$, $X$ is a new combination |
| **Step 3:** | Calculate the volume and weight of $X$ |
| **Step 4:** | If $X$ does not violate the limit $V$, $W$, put $X$ into $Comb$ and go back to **Step 2** |
| **Step 5:** | If $i = 0$, halt and all combination are found. |
| **Step 6:** | Let $x_i = lb_i$, $i = i - 1$，$X$ is a new combination，and go back to **Step 3** |

The improved multi-state BAT can check whether the state vector violates the limit (e.g., Step 4 in the pseudocode), so that the feasible state vectors can be added into the variable combination $Comb$. If a state vector does not meet the limit, it can be determined in advance that the next state vector obtained by the addition process must not meet the limit, so it can be stopped and move to another coordinate (as in Step 6 of the pseudocode). In this way, as the restrictions become more stringent, more steps can be skipped in the improved multi-state BAT, which can save more computation time. Therefore, this study will use the improved multi-state BAT to find all the combinations of integer variables more quickly.

## 3.3. Introduction of SSO and iSSO

SSO is an emerging population-based optimization algorithm first proposed by Yeh [24, 30]. It was initially designed to complement PSO in solving discrete problems, but eventually became a soft computing method with the advantages of simplicity, efficiency and flexibility. SSO has been applied in many fields [31-36].

Since there are two different types of variables in GRRAP, the BSSO proposed in this study will refer to two SSO algorithms that are suitable for different variables. These two types of SSO will be introduced and explained following. Let Nvar, Nsol, Ngen, and Nrun represent the number of variables, solutions, generations, and independent replications.

The traditional SSO is designed for discrete variables as in Eq. (10). Let $x_{ij}^t$, $P_{ij}^t$, and $g_j$ be the values of the $j$th variable in $X_i$, $P_i$, and $g$, respectively. $\rho_{ij}^t$ is a random number between 0 and 1. When $\rho_{ij}^t$ falls between 0 and $C_g$, $x_{ij}^{t+1}$ is updated according to the variable of the solution $gBest$, and when it falls between $C_g$ and $C_p$, $x_{ij}^{t+1}$ is updated according to the variable of the solution $pBest$. When it falls in the interval from $C_p$ to $C_w$, $x_{ij}^{t+1}$ does not make any change and directly takes the previous generation variable $x_{ij}^t$ as the new variable, and when it falls in the interval from $C_w$ to 1, $x_{ij}^{t+1}$ is newly generated between the upper and lower bounds of the variable

$$x_{ij}^{t+1} = \begin{cases} g_j & \text{if } \rho_{ij}^t \in [0, C_g = c_g) \\ P_{ij}^t & \text{if } \rho_{ij}^t \in [C_p, C_P = C_g + c_P) \\ x_{ij}^t & \text{if } \rho_{ij}^t \in [C_w, C_w = C_P + c_w) \\ x & \text{if } \rho_{ij}^t \in [C_w, 1) \end{cases} \quad (10)$$

$[-.5, .5]$ The iSSO is improved from the traditional SSO to solve continuous variables. It use $u_j$ in Eq. (12) to determine the search range and removing $pBest$, leaving only $gBest$ [37]. $\rho_{ij}^t$ is a random number between 0 and 1, and when $\rho_{ij}^t$ falls in the interval from 0 to $C_r$, $x_{ij}^{t+1}$ is updated to the neighbor of the $gBest$, and when it falls in the interval from $C_r$ to $C_g$, $x_{ij}^{t+1}$ is updated to the neighbor of the original variable $x_{ij}^t$, and when it falls between $C_g$ and 1, $x_{ij}^{t+1}$ is updated to a random range between the original variable $x_{ij}^t$ and $g_j$ as the new value. The iSSO is shown as Eq. (11) and Eq. (12).

## 4. Proposed BSSO

The proposed BSSO is formed by the combination of SSO, iSSO, multi-state BAT and network reliability methods. Referring to Huang [22], N-UM and R-UM are used to represent the update mechanisms of integer variables and continuous variables, respectively.

### 4.1. Solution representation

GRRAP needs to consider both the number of components and the reliability of the components. To make it easier to update them separately with different update mechanisms, the solution codes are divided into two segments according to the variable types. For example, in Fig.3, $X = (N, R)$ is $i$th solution, the first $N_{var}$ variables represent the number of components, and the last $N_{var}$ variables represent the reliability of the components.

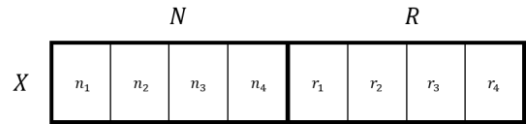

**Fig. 3.** Solution representation of $X = (N, R)$

$$x_{ij}^{t+1} = \begin{cases} x_{ij}^t + \rho_{[-.5,.5]} \cdot u_j & \text{if } x_{ij}^t = g_j \text{ or } \rho_{[0,1]} \in [0, C_r = c_r) \\ g_j + \rho \cdot u_j & \text{if } x_{ij}^t \neq g_j \text{ and } \rho_{[0,1]} \in [C_r, C_g = c_r + c_g) \\ x_{ij}^t + \rho_{[-.5,.5]} \cdot (x_{ij}^t - g_j) & \text{if } x_{ij}^t \neq g_j \text{ and } \rho_{[0,1]} \in [C_g, 1 = c_r + c_g + c_w] \end{cases} \quad (11)$$

where

$$u_j = \frac{x_j^{min} - x_j^{max}}{2 \cdot \text{Nvar}} \quad (12)$$

The major difference between BSSO and SSO is that BSSO recombines integer variables in advance by multi-state BAT and uses these combinations as the new integer variables. Since GRRAP has three main constraints, cost, size, and weight. The volume and weight are related to the number of components only, so it is possible to use multi-state BAT to find all the combinations of components that meet the volume and weight constraints and store them in the variable combinations set $Comb$. These combinations will be treated as new integer variable for SSO to select and update. This step also reduces the number of integer variable of the solution to one, thus reducing the dimension of the solution, as shown in Fig. 4. It also ensures that the solutions updated by SSO satisfy both the volume and weight constraints.

The complete process of multi-BAT filtering of integer variable combinations:

| Step 1: | BAT find all m-dimensional state vector $N_i = (n_1, n_2, ..., n_m)$ |
| Step 2: | Check whether the state vector $N_i$ meets the weight and volume limits, if not, go back to **Step 1** |
| Step 3: | Add the state vector $N_i$ that meet the constraints to the set of variable combination $Comb$. |

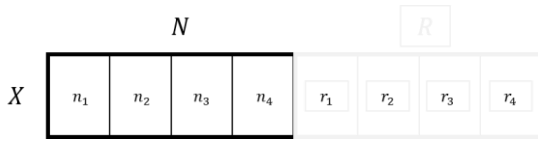

**Fig. 4 (a).** Original solution structure

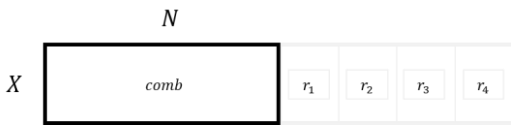

**Fig.4 (b).** Solution structure in BSSO

*4.2. N-UM for number of components*

The original integer update mechanism N-UM is the update mechanism of the traditional SSO, as shown in Eq. (13). The updated variable $n_{ij}^t$ is the number of components in each subsystem. BSSO has reorganized the integer variables using multi-state BAT, so the integer update mechanism N-UM is shown in Eq. (14). Therefore, the updated variable is changed from the number of components $n_{ij}^t$ to a feasible combination of the number of components $comb_i^t$. $\widehat{comb}_g$ is the variable of *gBest*, $\widehat{comb}_i$ is the variable of *pBest*.

$$n_{ij}^{t+1} = \begin{cases} \hat{n}_{g,j} & \text{if } \rho_{ij}^t \in [0, C_g = c_g) \\ \hat{n}_{i,j} & \text{if } \rho_{ij}^t \in [C_p, C_P = C_g + c_P) \\ n_{ij}^t & \text{if } \rho_{ij}^t \in [C_w, C_w = C_P + c_w) \\ n & \text{if } \rho_{ij}^t \in [C_w, 1) \end{cases} \quad (13)$$

$$comb_i^{t+1} = \begin{cases} \widehat{comb}_g & \text{if } \rho_i^t \in [0, C_g = c_g) \\ \widehat{comb}_i & \text{if } \rho_i^t \in [C_p, C_P = C_g + c_P) \\ comb_i^t & \text{if } \rho_i^t \in [C_w, C_w = C_P + c_w) \\ comb & \text{if } \rho_i^t \in [C_w, 1) \end{cases} \quad (14)$$

*4.3. R-UM for components reliability*

In the part of update mechanism R-UM, the update mechanism of the original iSSO is shown in Eq. (11). We believe that in RRAP, keeping the diversity of solutions will get better result, so the update mechanism reinserts *pBest* as the new R-UM shown in Eq. (15). $\hat{r}_{g,j}$ is the variable of *gBest*, $\tilde{r}_{i,j}$ is the variable of *pBest*.

$$r_{ij}^{t+1} = \begin{cases} \hat{r}_{g,j} + \rho_{[-.5,.5]} \cdot u_j & \text{if } r_{ij}^t \neq \hat{r}_{g,j} \text{ and } \rho_{[0,1]} \in [0, C_g = c_g) \\ \tilde{r}_{i,j} + \rho_{[-.5,.5]} \cdot u_j & \text{if } r_{ij}^t \neq \hat{r}_{g,j} \text{ and } \rho_{[0,1]} \in [C_p, C_p = C_g + c_p) \\ r_{ij}^t + \rho_{[-.5,.5]} \cdot u_j & \text{if } r_{ij}^t = \hat{r}_{g,j} \text{ or } \rho_{[0,1]} \in [C_p, C_w = C_p + c_w) \\ r_{ij}^t + \rho_{[-.5,.5]} \cdot (r_{ij}^t - \hat{r}_{g,j}) & \text{if } r_{ij}^t \neq \hat{r}_{g,j} \text{ and } \rho_{[0,1]} \in [C_w, 1] \end{cases} \quad (15)$$

where

$$u_j = \frac{r_j^{min} - r_j^{max}}{2(\frac{Ngen+t}{Ngen}) \cdot Nvar} \quad (16)$$

The parameter $u_j$ in iSSO is mainly used to determine the search range of variables, and it was a constant value in original iSSO. To make the search range vary in different generations and to enhance the local search ability in the later generations, the new $u_j$ is proposed and shown as Eq. (16).

If the updated variable $r_{ij}^{t+1}$ exceeds the upper and lower bounds, a new value within the upper and lower bounds is randomly generated as the value of $r_{ij}^{t+1}$.

### 4.3. The two-stage parameter $C_g$

The $C_g$ parameter in previous SSO studies is mostly constant in every generation [20, 32, 38], but in this study, the impact of *gBest* is adjusted by adjusting the $C_g$ parameter. The first half generations of $C_g$ is 0, and the original value is recovery after half the generations, which is a two-stage $C_g$, as in Eq. (17). Such an adjustment can preserve the diversity of solutions in the initial stage and improve the solution quality in the later stage by referring to the global best solution.

$$C_g = \begin{cases} 0 & \text{if } t < Ngen/2 \\ c_g & \text{otherwise} \end{cases} \quad (17)$$

### 4.4. Penalized reliability function and fitness function

Aiming at solving infeasible solutions that violate at least one constraint, the fitness function in the proposed BSSO is an adapted penalty reliability function that encourages the exploration of all possible solutions, both feasible and infeasible, in the solution space near the feasible solution bounds [11, 15, 22].

When a solution is obtained, the reliability of each subsystem in the system can be calculated, then the system reliability of whole GRRAP can be obtained by BAT network reliability method. $C_{ub}$, $V_{ub}$ and $W_{ub}$ are the upper bounds for the network cost, volume limit, and weight limit for the network, respectively. For a solution $X$ with a total system cost $g_c(N,R)$, volume $g_v(N,R)$, and weight $g_w(N,R)$, the fitness function is calculated that showed in Eq. (18).

$$F_R(X) = \begin{cases} R_s & \text{if } X = (N,R) \text{ is a feasible solution} \\ R_s \left( Min\left\{ \left[\frac{V_{ub}}{g_v(N,R)}\right], \left[\frac{C_{ub}}{g_c(N,R)}\right], \left[\frac{W_{ub}}{g_w(N,R)}\right] \right\} \right)^\gamma & \text{otherwise} \end{cases} \quad (18)$$

However, since BSSO already uses multi-state BAT to filter out integer combinations of variables that violate the volume and weight limits. Therefore, all the solutions updated by BSSO satisfy the volume and weight limits, which also leads to a penalty function that only needs to penalize solutions that violate the cost constraint. The fitness function in BSSO becomes as Eq. (19) shows.

$$F_R(X) = \begin{cases} R_s & \text{if } X = (N, R) \text{ is a feasible solution} \\ R_s \left( \dfrac{C_{ub}}{g_c(N,R)]} \right)^{\gamma} & \text{otherwise} \end{cases} \quad (19)$$

*4.5. Procedure of the proposed BSSO*

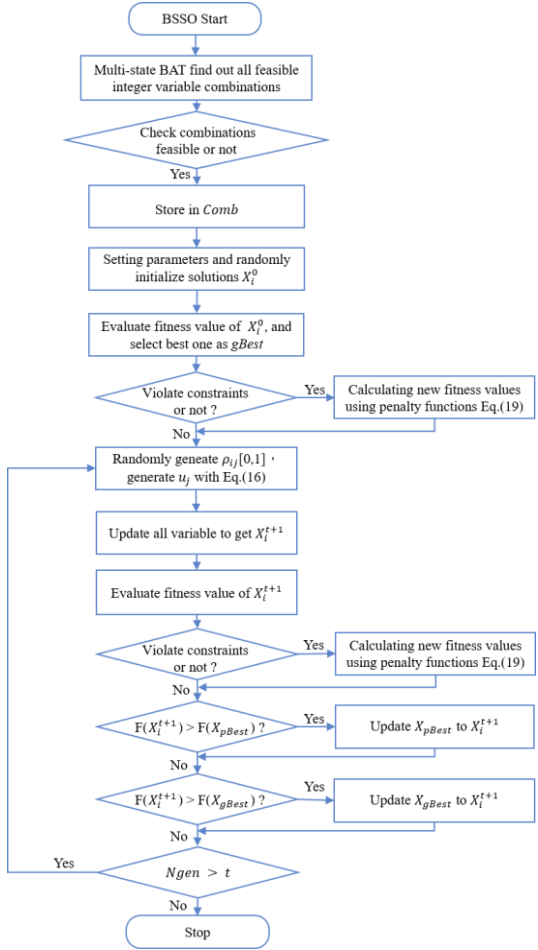

**Fig. 5.** Procedure of the proposed BSSO.

## 5. Numerical example

*5.1. Experimental setup*

The type of N-UM, whether the parameter $C_g$ changes, whether *pBest* is added back to R-UM, and whether $u_j$ in R-UM changes are the four factors incorporated in the proposed algorithm. Each of these four factors has two levels shown as Table 7, and the experimental design using $L_8(2^4)$ orthogonal array produced a total of eight combinations in Ex1 (Table 8). All tested algorithms were coded in Python version 3.8 and run on a desktop computer with AMD Ryzen 5 3600 processor, 3.59 GHz, 16 GB RAM, and Windows 10

For each tested algorithm, the number of solutions was set to 100 (Nsol = 100), the number of independent runs was set to 50, and the stopping criterion, i.e., the number of generators, was set to a maximum of 1000 (Ngen = 1000). Each solution in each algorithm was used to calculate the fitness function for that generation once. In addition, the fitness functions of all the algorithms are used in the method described in Section 3.1. The $\gamma$ in the Penalized reliability function is set to 3 with reference to the previous study [22].

All parameters required for GA, PSO, SSO, and nine combinations of Ex1 and Ex2 (including BSSO) are listed below:

BSSO: $C_g$ = 0.25, $C_p$ = 0.5, and $C_w$ = 0.6 (combination 8)

GA: Uniform mutation with a mutation rate of 0.4, two-point crossover with a crossover rate of 0.6, and elite selection

PSO: $w_{max} = 0.9, w_{min} = 0.4, c_1 = 2, c_2 = 2$, $w$ is linearly decreased from $w_{max}$ to $w_{min}$

SSO: $C_g$ = 0.25, $C_p$ = 0.5, and $C_w$ = 0.6 (combination 1)

Six benchmark problems in the field of network reliability were used to test all relevant algorithms, and all of them were adjusted to AON networks, as shown in Fig.7.

**Table 7** Factor level table

| Factor | Level 1 | Level 2 |
|---|---|---|
| N-UM | $n_{ij}^{t+1} = \begin{cases} \hat{n}_{g,j} & \text{if } \rho_{ij}^t \in [0, C_g) \\ \hat{n}_{i,j} & \text{if } \rho_{ij}^t \in [C_p, C_P) \\ n_{ij}^t & \text{if } \rho_{ij}^t \in [C_w, C_w) \\ n & \text{if } \rho_{ij}^t \in [C_w, 1) \end{cases}$ | $comb_i^{t+1} = \begin{cases} \widehat{comb}_g & \text{if } \rho_i^t \in [0, C_g) \\ \widehat{comb}_i & \text{if } \rho_i^t \in [C_p, C_P) \\ comb_i^t & \text{if } \rho_i^t \in [C_w, C_w) \\ comb & \text{if } \rho_i^t \in [C_w, 1) \end{cases}$ |
| $C_g$ | $C_g = c_g$ | $C_g = \begin{cases} 0 & \text{if } t < \text{Ngen}/2 \\ c_g & \text{otherwise} \end{cases}$ |
| R-UM | $r_{ij}^{t+1} = \begin{cases} \hat{r}_{g,j} + \rho_{[-.5,.5]} \cdot u_j & \text{if } r_{ij}^t \neq \hat{r}_{g,j} \text{ and } \rho_{[0,1]} \\ r_{ij}^t + \rho_{[-.5,.5]} \cdot u_j & \text{if } r_{ij}^t = \hat{r}_{g,j} \text{ or } \rho_{[0,1]} \\ r_{ij}^t + \rho_{[-.5,.5]} \cdot (r_{ij}^t - \hat{r}_{g,j}) & \text{if } r_{ij}^t \neq g_j \text{ and } \rho_{[0,1]} \end{cases}$ | $r_{ij}^{t+1} = \begin{cases} \hat{r}_{g,j} + \rho_{[-.5,.5]} \cdot u_j & \text{if } r_{ij}^t \neq \hat{r}_{g,j} \text{ and } \rho_{[0,1]} \\ \tilde{r}_{i,j} + \rho_{[-.5,.5]} \cdot u_j & \text{if } r_{ij}^t \neq \hat{r}_{g,j} \text{ and } \rho_{[0,1]} \\ r_{ij}^t + \rho_{[-.5,.5]} \cdot u_j & \text{if } r_{ij}^t = \hat{r}_{g,j} \text{ or } \rho_{[0,1]} \\ r_{ij}^t + \rho_{[-.5,.5]} \cdot (r_{ij}^t - \hat{r}_{g,j}) & \text{if } r_{ij}^t \neq \hat{r}_{g,j} \text{ and } \rho_{[0,1]} \end{cases}$ |
| $u_j$ | $u_j = \dfrac{r_j^{min} - r_j^{max}}{2 \cdot \text{Nvar}}$ | $u_j = \dfrac{r_j^{min} - r_j^{max}}{2(\dfrac{\text{Ngen}+t}{\text{Ngen}}) \cdot \text{Nvar}}$ |

The first four benchmark problems are the same as those used in GRAP [15], and two additional larger networks are added [23]. It is worth noting that the first four benchmark problems are used in Ex1 to test the performance of the algorithm combinations, while Ex2 uses all benchmark problems for the comparison of the algorithms.

In all six benchmark problems, the reliability, cost, volume, and weight of each component are based on the data from the RRAP research of Hikita et al. and Dhingra [14, 39]. Detailed data are listed in the Appendix. In addition, to make the results more distinguishable, adjustments will be made for the three limitations of cost, volume, and weight. Let the notation $F_{avg}$, $F_{max}$, $F_{min}$, and $F_{stdev}$, indicate the mean, maximum (best), minimum (worst), and standard deviation, respectively. Let $T_{avg}$ be the average running time. In addition, the bold values denote the best results of the test problem.

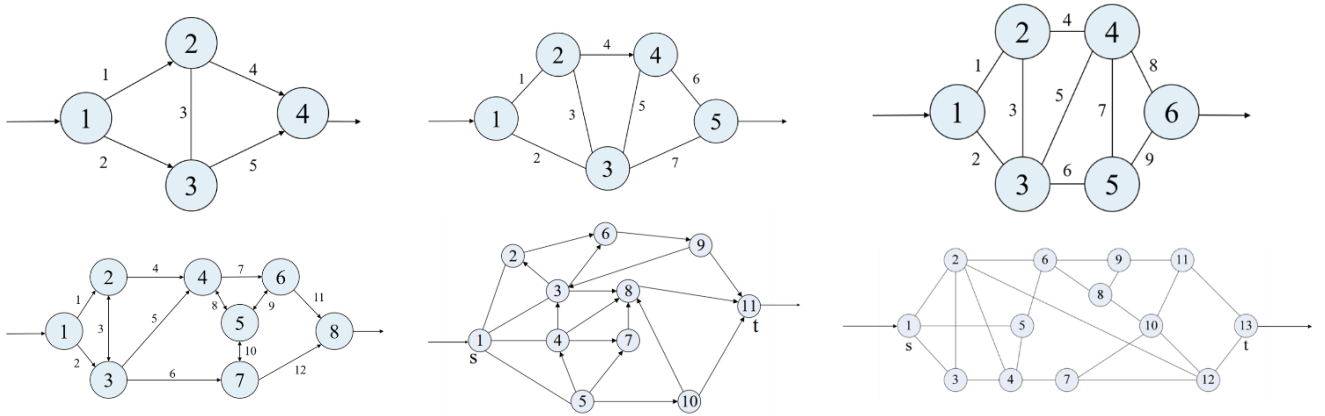

**Fig. 6.** Network diagram of the six benchmark problems.

*5.2. Ex1: finding the best combination*

The result of the 50 independent experiments for each of the 8 combinations is shown in Table 9-14. The values in the table are the fitness values obtained by the algorithm.

**Table 9** Overall results of eight combinations for BSSO in benchmark 1.

| No. | $F_{avg}$ | $F_{stdev}$ | $F_{max}$ | $F_{min}$ |
|---|---|---|---|---|
| 1 | 0.9758844807 | 0.0013448597 | 0.9765968718 | 0.9712918309 |
| 2 | 0.9758797390 | 0.0016140715 | 0.9766026878 | 0.9714137324 |
| 3 | 0.9758847098 | 0.0015017933 | 0.9766216746 | 0.9713328318 |
| 4 | 0.9762916136 | 0.0002258091 | 0.9766010825 | 0.9752591320 |
| 5 | 0.9765582325 | 0.0000517135 | **0.9766346548** | 0.9763452985 |
| 6 | 0.9765221505 | 0.0000567250 | 0.9766289410 | 0.9764161625 |
| 7 | 0.9764037829 | 0.0001493598 | 0.9766303825 | 0.9758557315 |
| 8 | **0.9765765892** | **0.0000296157** | 0.9766322743 | **0.9765144736** |
| 9 | 0.9764926010 | 0.0000769853 | 0.9766205061 | 0.9763310255 |

**Table 10** Overall results of eight combinations for BSSO in benchmark 2.

| No. | $F_{avg}$ | $F_{stdev}$ | $F_{max}$ | $F_{min}$ |
|---|---|---|---|---|
| 1 | 0.9950529947 | 0.0002522171 | 0.9953591780 | 0.9945261946 |
| 2 | 0.9952300200 | 0.0001442508 | 0.9953623098 | 0.9947635892 |
| 3 | 0.9951118221 | 0.0002758800 | 0.9953634134 | 0.9939635610 |
| 4 | 0.9952156178 | 0.0000927362 | 0.9953817582 | 0.9949017554 |
| 5 | 0.9952539071 | 0.0002195056 | **0.9954502752** | 0.9947238361 |
| 6 | 0.9952611711 | 0.0001761174 | 0.9954424849 | 0.9947251596 |
| 7 | 0.9952629972 | 0.0001303164 | 0.9953937764 | 0.9948250707 |
| 8 | **0.9953583757** | **0.0000315751** | 0.9954298077 | **0.9952747159** |

**Table 11** Overall results of eight combinations for BSSO in benchmark 3.

| No. | $F_{avg}$ | $F_{stdev}$ | $F_{max}$ | $F_{min}$ |
|---|---|---|---|---|
| 1 | 0.9954575001 | 0.0005549384 | 0.9963741061 | 0.9933866372 |
| 2 | 0.9957516843 | 0.0004278140 | 0.9964189726 | 0.9942283490 |
| 3 | 0.9956640209 | 0.0002935374 | 0.9963832146 | 0.9950776347 |
| 4 | 0.9955978581 | 0.0002998385 | 0.9963022427 | 0.9950291115 |
| 5 | 0.9958348099 | 0.0003981162 | 0.9964173750 | 0.9952408678 |
| 6 | 0.9958385357 | 0.0003676733 | 0.9964105251 | 0.9951445707 |
| 7 | 0.9957038251 | 0.0003623215 | 0.9963950593 | 0.9950261873 |
| 8 | **0.9963297279** | **0.0001770982** | **0.9964272846** | **0.9956984483** |

**Table 12** Overall results of eight combinations for BSSO in benchmark 4.

| No. | $F_{avg}$ | $F_{stdev}$ | $F_{max}$ | $F_{min}$ |
|---|---|---|---|---|
| 1 | 0.9983664768 | 0.0009099681 | 0.9989740925 | 0.9937572894 |
| 2 | 0.9987179769 | 0.0002807252 | 0.9990856572 | 0.9978838841 |
| 3 | 0.9984959950 | 0.0004580234 | 0.9990787505 | 0.9971666361 |
| 4 | 0.9986088023 | 0.0003206662 | 0.9989868418 | 0.9975749729 |
| 5 | 0.9987356853 | 0.0003618973 | 0.9991471500 | 0.9977291209 |
| 6 | 0.9988241703 | 0.0002615493 | 0.9991321419 | 0.9979941116 |
| 7 | 0.9986470691 | 0.0002923474 | 0.9991002030 | 0.9979033734 |
| 8 | **0.9990157953** | **0.0000877533** | **0.9991517465** | **0.9988319879** |

From the experimental results, it is observed that combination 8 has the best average value for all benchmark problems. For the maximum values, the best results are obtained in benchmark 3 and 4 and the average results. Overall, the algorithm of combination 8 (all factors set at level 2) has the best performance in the first four benchmark problems.

The above observations were further investigated using statistical analysis to find the effect of each factor on the response values. Analysis of Variance, main effects analysis and interaction analysis will be performed. Through these results, the optimal combination of factor levels will be identified. The following benchmark problem 4 is taken as an illustration, and the rest of the detailed results can be seen in the Appendix.

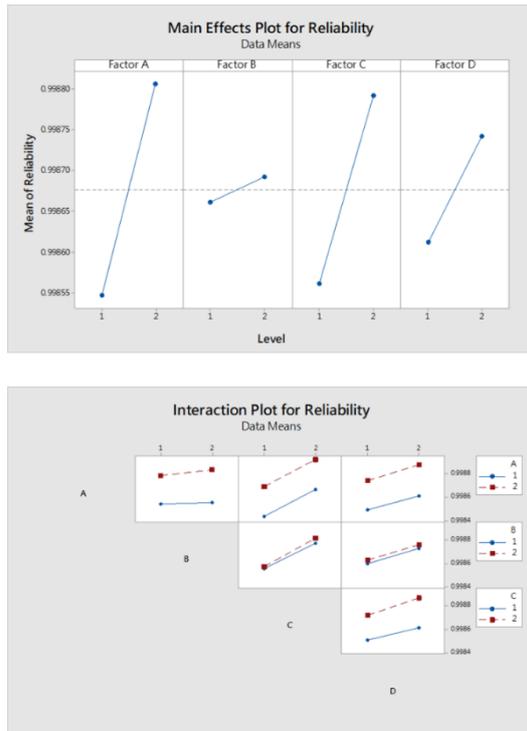

**Fig. 4.** Main effects analysis and interaction analysis for reliability of benchmark 4.

**Table 4** ANOVA table for reliability of benchmark 4.

| Source | DF | SS | MS | F-value | P-value |
|---|---|---|---|---|---|
| A | 1 | 0.000000 | 0.000000 | 372.14 | 0.000*** |
| B | 1 | 0.000000 | 0.000000 | 5.30 | 0.105 |
| C | 1 | 0.000000 | 0.000000 | 295.88 | 0.000*** |
| D | 1 | 0.000000 | 0.000000 | 93.83 | 0.002** |
| Error | 3 | 0.000000 | 0.000000 | | |
| Total | 7 | 0.000000 | 0.000000 | | |
| S=0.00001894 | | R-Sq=99.6% | | R-Sq(adj)=99.1% | |

* P≤0.05, ** P≤0.01, *** P≤0.001

From the main effects and interaction analysis in Fig. 8, we can observe that all the factors have better responses when they are set at level 2. The ANOVA in Table 15 shows that the p-values of the three factors A, C, and D are less than 0.05, reject the null hypothesis and showing a significant effect on the response values when the confidence level is 95%.

Based on the above analysis, we can see that the way of finding all feasible combinations of integer variables with multi-state BAT (Factor A) has a significant impact on the improvement of the solution results. Moreover, since combination 8 is the best one, it will also be the final version of BSSO.

### 5.3. Ex2: comparing the GA, PSO, SSO, and BSSO

#### 5.3.1. Solution comparison

It can be observed from Table 14 that the proposed BSSO has the best performance in all problems in terms of $F_{avg}$, $F_{min}$, and $F_{stdev}$ of the fitness values. However, in terms of the $F_{max}$, although PSO performs better in benchmark 1-3, the difference with BSSO is very small. But as the size of the problem increases, $F_{max}$ of BSSO outperforms the algorithms in benchmark 4-6. Since PSO is better at updating continuous variables, the disadvantage of PSO in updating integer variables is not obvious when the problem size is small. When the problem size increases, the disadvantage becomes obvious, resulting in worse results than the other algorithms. In terms of overall results, BSSO is the best performing algorithm in terms of solution quality, obtaining higher quality solutions for each problem of different sizes.

According to Table 14, the $T_{avg}$ of BSSO is the shortest in problems 1-4. The reason is that BSSO has used multi-state BAT to find out the solutions that satisfy the weight and volume limits. Therefore, there is no need to recalculate and check the volume and weight of the solutions during the iterative process. The rest of the algorithms require a lot of volume and weight calculations during the iterative process to be applied to the penalized fitness function. However, in problems 5-6, as the problem expands, the time required to find all feasible combinations of variables in the multi-state BAT increases significantly, resulting in a higher $T_{avg}$ for the BSSO than the other algorithms. Based on the result, it can be found that even for larger problems, if the problem weight and volume restrictions are strict enough. Even if BSSO additionally uses multi-state BAT to find out all feasible integer variable combinations, it does not take too much computation time, making the overall computation time is similar or even faster than other algorithms.

**Table 14** BSSO, GA, PSO, and SSO results.

| ID | method | $F_{avg}$ | $F_{max}$ | $F_{min}$ | $F_{stdev}$ | $T_{avg}$ |
|---|---|---|---|---|---|---|
| 1 | BSSO | **0.976577** | 0.976646 | **0.976462** | **0.000038** | **2.307349** |
|   | GA   | 0.970359 | 0.976507 | 0.957192 | 0.005123 | 3.924821 |
|   | PSO  | 0.974756 | **0.976649** | 0.971617 | 0.002402 | 3.499959 |
|   | SSO  | 0.975762 | 0.976550 | 0.971143 | 0.001480 | 2.819058 |
| 2 | BSSO | **0.995362** | 0.995430 | **0.995258** | **0.000036** | **2.925581** |
|   | GA   | 0.993871 | 0.995250 | 0.988281 | 0.001643 | 3.879048 |
|   | PSO  | 0.995025 | **0.995483** | 0.991496 | 0.000916 | 4.418789 |
|   | SSO  | 0.995067 | 0.995362 | 0.994173 | 0.000271 | 3.669631 |
| 3 | BSSO | **0.996283** | 0.996423 | **0.995628** | **0.000243** | **3.628899** |
|   | GA   | 0.993580 | 0.996165 | 0.988294 | 0.002173 | 4.690570 |
|   | PSO  | 0.995660 | **0.996430** | 0.992077 | 0.000611 | 5.399903 |
|   | SSO  | 0.995379 | 0.996418 | 0.993021 | 0.000615 | 4.530452 |
| 4 | BSSO | **0.999025** | **0.999152** | **0.998839** | **0.000096** | **7.105688** |
|   | GA   | 0.997749 | 0.999027 | 0.990882 | 0.001337 | 7.873402 |
|   | PSO  | 0.997568 | 0.999019 | 0.993861 | 0.001351 | 8.975268 |
|   | SSO  | 0.998540 | 0.999118 | 0.996217 | 0.000545 | 7.831062 |
| 5 | BSSO | **0.995358** | **0.995847** | **0.994120** | **0.000406** | 57.682561 |
|   | GA   | 0.985215 | 0.995002 | 0.938137 | 0.009893 | 57.268283 |
|   | PSO  | 0.956265 | 0.987118 | 0.845108 | 0.029860 | 59.371770 |
|   | SSO  | 0.986015 | 0.994312 | 0.970786 | 0.006025 | **56.580192** |
| 6 | BSSO | **0.987793** | **0.988890** | **0.985012** | **0.000826** | 155.586186 |
|   | GA   | 0.966887 | 0.987143 | 0.886465 | 0.016875 | 154.577601 |
|   | PSO  | 0.761224 | 0.952302 | 0.361390 | 0.142731 | 156.991670 |
|   | SSO  | 0.930147 | 0.982299 | 0.821940 | 0.043094 | **152.319698** |
| Avg | BSSO | **0.991733** | **0.992065** | **0.990886** | **0.000274** | 38.206044 |
|     | GA   | 0.984610 | 0.991516 | 0.958208 | 0.006174 | 38.702287 |
|     | PSO  | 0.946750 | 0.984500 | 0.859258 | 0.029645 | 39.776227 |
|     | SSO  | 0.980152 | 0.990676 | 0.957880 | 0.008672 | **37.958349** |

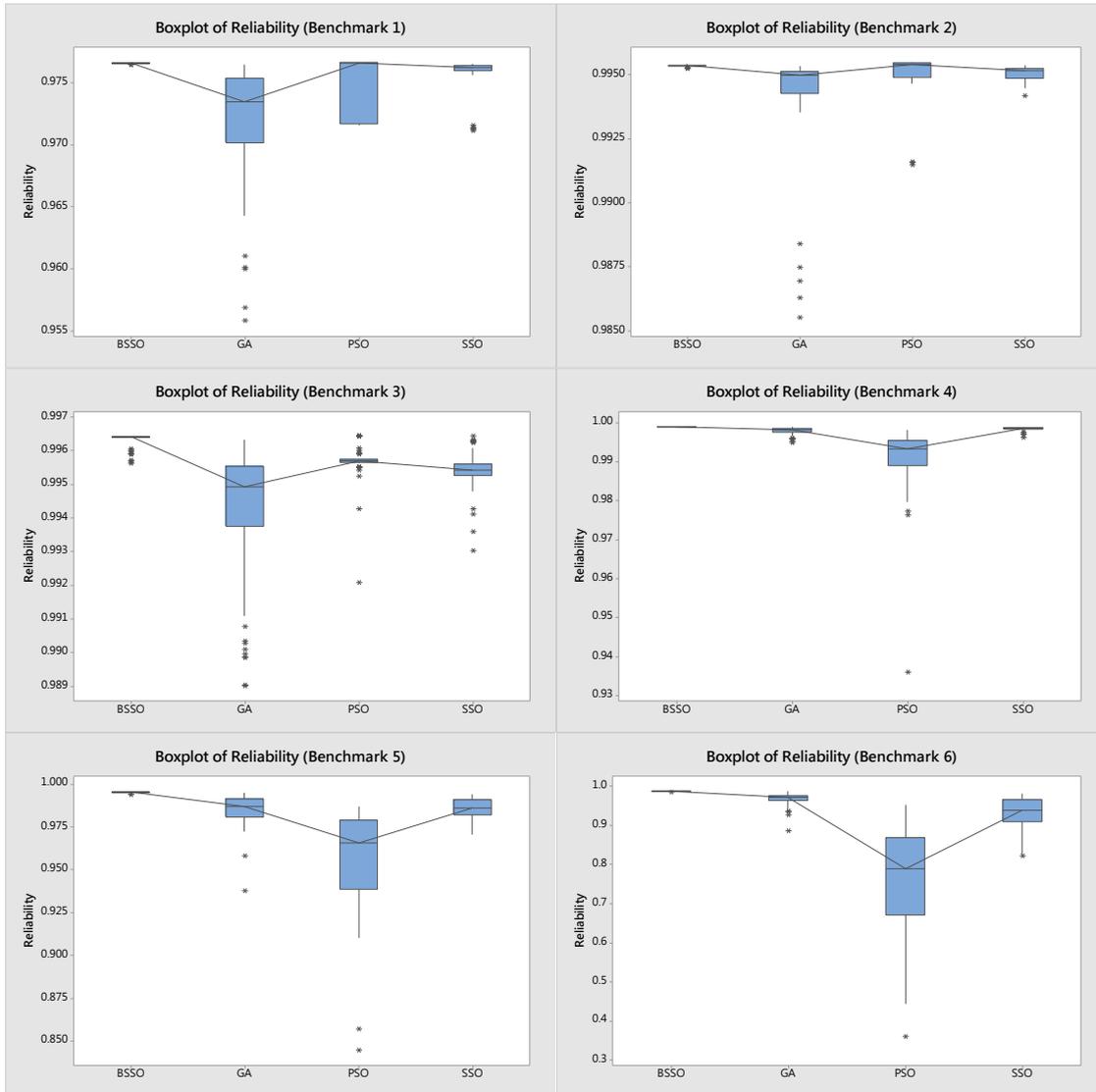

**Fig.8.** Boxplot for the reliability in Ex2.

Based on Table 14, it is observed that BSSO has the smallest $F_{stdev}$ in all benchmarks. This means that BSSO has been able to obtain stable results in several independent experiments. Also, the standard deviation of BSSO does not significantly change when facing different problems because of the difference in the number of nodes and network structure. The boxplot in Fig.8 also shows that the distribution of BSSO fitness values is highly consistent across all benchmark problems, and there are few outliers. It can be seen that BSSO performs much better than the other three algorithms (GA, PSO, and SSO) in terms of stability

## 6. Conclusion

In this study, a GRRAP problem and a new soft computing method BSSO are proposed. The proposed BSSO is a combination of multi-state BAT and SSO with iSSO, and some update mechanisms are adjusted for GRRAP. The BAT method is then used for network reliability calculation. The traditional RRAP is a common reliability design problem. The proposed GRRAP is an extension of RRAP, which allows the system configuration to be extended to general network structure without the limit of series or series-parallel connection, making the application scope more extensive.

The proposed BSSO performs the best in the comparison experiments with GA, PSO, and SSO, and not only obtains high quality solutions but also has high stability. From the experimental results, BSSO is a promising soft computing method that combines the efficiency of SSO with the accuracy of BAT.

The proposed GRRAP is an NP-hard problem, especially the network reliability is very difficult to solve. Even though the BAT method is one of the most efficient methods to calculate the binary-state network reliability, it still requires a very large amount of computation time when the network system is scaled up. Therefore, in the future, it may be possible to use Monte Carlo simulation method to obtain approximate values of network reliability instead of exact values to reduce the computation burden [40]. In addition, the proposed GRRAP can be extended to more types of redundant strategies such as cold-standby and mixed-standby. Ultimately, BSSO is expected to be applied to problems in different fields.

## APPENDIX

**Table 1** Data from the six GRRAP test problems.

| ID | $i$ | $10^5 \alpha_i$ | $\beta_i$ | $w_i v_i^2$ | $w_i$ | $V$ | $C$ | $W$ |
|---|---|---|---|---|---|---|---|---|
| 1 | 1 | 1.0 | 1.5 | 1 | 6 | 50 | 80 | 100 |
| | 2 | 2.3 | 1.5 | 2 | 6 | | | |
| | 3 | 0.3 | 1.5 | 3 | 8 | | | |
| | 4 | 2.3 | 1.5 | 2 | 7 | | | |
| 2 | 1 | 2.330 | 1.5 | 1 | 7 | 110 | 175 | 200 |
| | 2 | 1.450 | 1.5 | 2 | 8 | | | |
| | 3 | 0.541 | 1.5 | 3 | 8 | | | |
| | 4 | 8.050 | 1.5 | 4 | 6 | | | |
| | 5 | 1.950 | 1.5 | 2 | 9 | | | |
| 3 and 4 | 1 | 2.5 | 1.5 | 2 | 3.5 | 220 | 210 | 120 |
| | 2 | 1.45 | 1.5 | 4 | 4.0 | 285 | 280 | 160 |
| | 3 | 0.541 | 1.5 | 5 | 4.0 | | | |
| | 4 | 0.541 | 1.5 | 8 | 3.5 | | | |
| | 5 | 2.1 | 1.5 | 4 | 4.5 | | | |
| | 6 | 2.1 | 1.5 | 4 | 4.5 | | | |
| | 7 | 1.45 | 1.5 | 4 | 4.0 | | | |
| | 8 | 0.541 | 1.5 | 5 | 4.0 | | | |
| 5 and 6 | 1 | 2.5 | 1.5 | 2 | 3.5 | 225 | 225 | 130 |
| | 2 | 1.45 | 1.5 | 4 | 4.0 | 225 | 225 | 130 |
| | 3 | 0.541 | 1.5 | 5 | 4.0 | | | |
| | 4 | 0.541 | 1.5 | 8 | 3.5 | | | |
| | 5 | 2.1 | 1.5 | 4 | 4.5 | | | |
| | 6 | 2.5 | 1.5 | 2 | 3.5 | | | |
| | 7 | 1.45 | 1.5 | 4 | 4.0 | | | |
| | 8 | 0.541 | 1.5 | 5 | 4.0 | | | |
| | 9 | 0.541 | 1.5 | 8 | 3.5 | | | |
| | 10 | 2.1 | 1.5 | 4 | 4.5 | | | |
| | 11 | 2.5 | 1.5 | 2 | 3.5 | | | |
| | 12 | 1.45 | 1.5 | 4 | 4.0 | | | |
| | 13 | 0.541 | 1.5 | 5 | 4.0 | | | |

Statistical analysis of benchmark problem 1

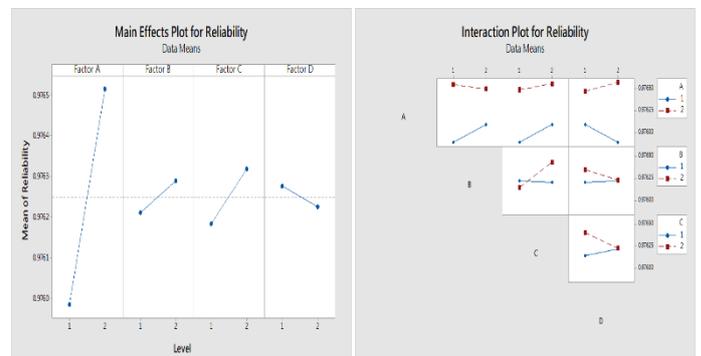

| Source | DF | SS | MS | F-value | P-value |
|---|---|---|---|---|---|
| A | 1 | 0.000000 | 0.000000 | 372.14 | 0.000*** |

| | 1 | 0.00 | 0.00 | 5.30 | 0.105 |
|---|---|---|---|---|---|
| B | | 0000 | 0000 | | |
| C | 1 | 0.00 0000 | 0.00 0000 | 295.88 | 0.000*** |
| D | 1 | 0.00 0000 | 0.00 0000 | 93.83 | 0.002** |
| Error | 3 | 0.00 0000 | 0.00 0000 | | |
| Total | 7 | 0.00 0000 | 0.00 0000 | | |
| S=0.00001894 | | R-Sq=99.6% | | R-Sq (adj)= 99.1% | |

* P≤0.05, ** P≤0.01, *** P≤0.001

Statistical analysis of benchmark problem 2

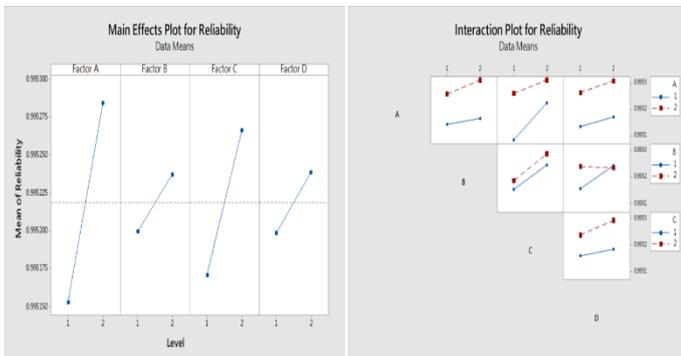

| Source | DF | SS | MS | F-value | P-value |
|---|---|---|---|---|---|
| A | 1 | 0.00 0000 | 0.00 0000 | 372.14 | 0.000*** |
| B | 1 | 0.00 0000 | 0.00 0000 | 5.30 | 0.105 |
| C | 1 | 0.00 0000 | 0.00 0000 | 295.88 | 0.000*** |
| D | 1 | 0.00 0000 | 0.00 0000 | 93.83 | 0.002** |
| Error | 3 | 0.00 0000 | 0.00 0000 | | |
| Total | 7 | 0.00 0000 | 0.00 0000 | | |
| S=0.00001894 | | R-Sq=99.6% | | R-Sq (adj)= 99.1% | |

Statistical analysis of benchmark problem 3

| Source | DF | SS | MS | F-value | P-value |
|---|---|---|---|---|---|
| A | 1 | 0.00 0000 | 0.00 0000 | 372.14 | 0.000*** |
| B | 1 | 0.00 0000 | 0.00 0000 | 5.30 | 0.105 |
| C | 1 | 0.00 0000 | 0.00 0000 | 295.88 | 0.000*** |
| D | 1 | 0.00 0000 | 0.00 0000 | 93.83 | 0.002** |
| Error | 3 | 0.00 0000 | 0.00 0000 | | |
| Total | 7 | 0.00 0000 | 0.00 0000 | | |
| S=0.00001894 | | R-Sq=99.6% | | R-Sq (adj)= 99.1% | |

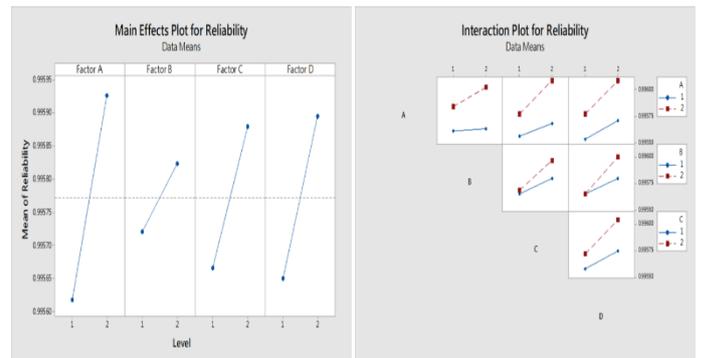

**Table 2** The final results obtained from BSSO, GA, PSO, and SSO in benchmark 1

| | BSSO | GA | PSO | SSO |
|---|---|---|---|---|
| $n$ | (2, 1, 2, 3) | (2, 1, 2, 3) | (2, 1, 2, 3) | (2, 1, 2, 3) |
| $r_1$ | 0.894528 | 0.893314 | 0.894598 | 0.891724 |
| $r_2$ | 0.583716 | 0.526231 | 0.584160 | 0.602920 |
| $r_3$ | 0.904896 | 0.913404 | 0.904167 | 0.899354 |
| $r_4$ | 0.794840 | 0.794101 | 0.795176 | 0.801045 |
| $R$ | 0.9766455382 | 0.9765070256 | **0.9766489194** | 0.9765497536 |

**Table 3** The final results obtained from BSSO, GA, PSO, and SSO in benchmark 2

| | BSSO | GA | PSO | SSO |
|---|---|---|---|---|
| $n$ | (3, 2, 3, 1, 3) | (3, 1, 3, 2, 3) | (3, 1, 3, 1, 3) | (3, 1, 3, 2, 3) |
| $r_1$ | 0.874185 | 0.865531 | 0.874880 | 0.873016 |
| $r_2$ | 0.209394 | 0.559355 | 0.107096 | 0.339325 |
| $r_3$ | 0.905545 | 0.900467 | 0.906250 | 0.901538 |
| $r_4$ | 0.0512 | 0.1568 | 0.0034 | 0.1490 |
|   | 40 | 75 | 64 | 29 |
| $r_5$ | 0.8793 03 | 0.8855 46 | 0.8797 00 | 0.8798 99 |
| $R$ | 0.9954 302822 | 0.9952 496200 | **0.9954 830911** | 0.9953 616775 |

**Table 4** The final results obtained from BSSO, GA, PSO, and SSO in benchmark 3

|   | BSSO | GA | PSO | SSO |
|---|---|---|---|---|
| $n$ | (4, 2, 2, 2, 2, 3) | (4, 2, 2, 2, 2, 3) | (4, 2, 2, 2, 2, 3) | (4, 2, 2, 2, 2, 3) |
| $r_1$ | 0.8222 30 | 0.8264 97 | 0.8239 81 | 0.8253 10 |
| $r_2$ | 0.7912 77 | 0.7701 10 | 0.7859 51 | 0.7903 78 |
| $r_3$ | 0.8992 35 | 0.8925 43 | 0.9006 24 | 0.9006 73 |
| $r_4$ | 0.9093 21 | 0.9217 81 | 0.9062 48 | 0.9073 59 |
| $r_5$ | 0.7522 24 | 0.7926 19 | 0.7556 41 | 0.7598 26 |
| $r_6$ | 0.8839 95 | 0.8740 94 | 0.8840 28 | 0.8818 32 |
| $R$ | 0.9964 225431 | 0.9961 651960 | **0.9964 295536** | 0.9964 181892 |

**Table 5** The final results obtained from BSSO, GA, PSO, and SSO in benchmark 4

|   | BSSO | GA | PSO | SSO |
|---|---|---|---|---|
| $n$ | (4, 1, 3, 2, 1, 3, 2, 3) | (4, 1, 3, 2, 1, 3, 2, 3) | (4, 2, 3, 3, 1, 2, 2, 3) | (4, 1, 3, 2, 1, 3, 2, 3) |
| $r_1$ | 0.8691 02 | 0.8815 54 | 0.8636 42 | 0.8668 16 |
| $r_2$ | 0.7474 83 | 0.8440 46 | 0.7381 19 | 0.7561 88 |
| $r_3$ | 0.9189 43 | 0.9093 95 | 0.8918 12 | 0.9179 46 |
| $r_4$ | 0.8675 83 | 0.8783 73 | 0.8358 01 | 0.8306 82 |
| $r_5$ | 0.0455 60 | 0.4128 48 | 0.0072 47 | 0.3959 14 |
| $r_6$ | 0.8430 24 | 0.8188 31 | 0.8719 62 | 0.8617 46 |
| $r_7$ | 0.8263 62 | 0.8316 78 | 0.8775 87 | 0.7887 41 |
| $r_8$ | 0.9396 76 | 0.9271 40 | 0.9406 07 | 0.9396 36 |
| $R$ | **0.9991 521239** | 0.9990 273441 | 0.9990 188162 | 0.9991 178284 |

**Table 6** The final results obtained from BSSO, GA, PSO, and SSO in benchmark 5

|   | BSSO | GA | PSO | SSO |
|---|---|---|---|---|
| $n$ | (4, 1, 2, 2, 1, 1, 1, 2, 1, 1, 3) | (4, 1, 2, 1, 1, 1, 1, 2, 2, 1, 3) | (3, 1, 1, 3, 1, 2, 1, 2, 2, 1, 3) | (4, 1, 2, 2, 1, 1, 1, 2, 1, 1, 3) |
| $r_1$ | 0.8194 61 | 0.8261 85 | 0.8081 99 | 0.8175 91 |
| $r_2$ | 0.3284 71 | 0.7082 76 | 0.5206 56 | 0.8819 20 |
| $r_3$ | 0.8396 79 | 0.9179 17 | 0.8504 98 | 0.8145 36 |
| $r_4$ | 0.8320 91 | 0.8048 05 | 0.9013 71 | 0.8359 15 |
| $r_5$ | 0.7289 13 | 0.7565 13 | 0.3395 34 | 0.7688 97 |
| $r_6$ | 0.7016 51 | 0.6332 88 | 0.6183 81 | 0.5870 87 |
| $r_7$ | 0.2126 58 | 0.2617 01 | 0.5491 34 | 0.3102 95 |
| $r_8$ | 0.9293 99 | 0.9164 31 | 0.8738 75 | 0.9228 94 |
| $r_9$ | 0.8268 25 | 0.6250 82 | 0.6868 49 | 0.7097 09 |
| $r_{10}$ | 0.7186 45 | 0.6602 09 | 0.3275 90 | 0.6195 31 |
| $r_{11}$ | 0.8778 40 | 0.8709 55 | 0.8786 76 | 0.8639 55 |
| $R$ | **0.9958 472260** | 0.9950 017722 | 0.9871 182661 | 0.9943 118982 |

**Table 7** The final results obtained from BSSO, GA, PSO, and SSO in benchmark 6

|   | BSSO | GA | PSO | SSO |
|---|---|---|---|---|
| $n$ | (3, 2, 1, 1, 1, 1, 1, 1, 1, 1, 2, 2, 3) | (3, 2, 1, 1, 1, 1, 1, 1, 1, 1, 2, 3, 2) | (3, 2, 1, 2, 1, 1, 1, 2, 1, 2, 1, 1, 2) | (3, 2, 1, 1, 1, 1, 2, 1, 1, 1, 1, 3, 2) |
| $r_1$ | 0.8508 77 | 0.8256 79 | 0.8014 71 | 0.8246 37 |
| $r_2$ | 0.8802 93 | 0.8832 52 | 0.8747 90 | 0.8943 84 |
| $r_3$ | 0.7960 28 | 0.7662 02 | 0.5103 77 | 0.5925 95 |
| $r_4$ | 0.8562 32 | 0.8677 86 | 0.6703 00 | 0.5465 53 |
| $r_5$ | 0.7239 35 | 0.7638 54 | 0.7682 64 | 0.6811 09 |
| $r_6$ | 0.7271 90 | 0.7277 57 | 0.5243 62 | 0.8282 41 |
| $r_7$ | 0.8131 93 | 0.8547 28 | 0.8061 82 | 0.7426 63 |
| $r_8$ | 0.5829 16 | 0.6556 42 | 0.2004 53 | 0.7634 98 |
| $r_9$ | 0.8232 78 | 0.7177 25 | 0.5446 11 | 0.3657 07 |
| $r_{10}$ | 0.658254 | 0.719660 | 0.769257 | 0.602330 |
| $r_{11}$ | 0.672413 | 0.597652 | 0.830824 | 0.414248 |
| $r_{12}$ | 0.887437 | 0.837189 | 0.932718 | 0.885617 |
| $r_{13}$ | 0.893725 | 0.939390 | 0.884480 | 0.917633 |
| $R$ | **0.9888899307** | 0.9871433406 | 0.9523024807 | 0.9822988387 |